\documentclass{article}

\usepackage{graphicx}
\usepackage{subfig}
\usepackage{multirow}
\usepackage{cases}
\usepackage{epsfig}
\usepackage{a4,cite}

\newcommand{\comment}[1]{}
\newcommand{\Paragraph}[1]{\smallskip\noindent {\bf #1}.\ }

\begin{document}
\begin{center}

\vspace*{1.0in}

\huge{\bf Energy-Efficient Design and Optimization of Wireline Access Networks}

\vspace*{1.0in}

\begin{table}[!h]
\begin{tabular}{cccc}
\textbf{Sourjya Bhaumik} & \textbf{David Chuck} & \textbf{Girija Narlikar} & \textbf{Gordon Wilfong} \\
Bell Labs India & Iowa State University & Bell Labs India & Bell Labs USA \\
Alcatel-Lucent & Ames, Iowa & Alcatel-Lucent & Alcatel-Lucent \\
\end{tabular}
\end{table}

\vspace*{3.0in}

\Large{\bf Technical Report}\\
\large{\bf January 2011}

\end{center}

\comment{
\title{Energy-Efficient Design and Optimization of Wireline Access Networks}

\author{\IEEEauthorblockN{Sourjya Bhaumik}
\IEEEauthorblockA{Bell Labs India\\
Alcatel-Lucent\\
}
\and
\IEEEauthorblockN{David Chuck}
\IEEEauthorblockA{Iowa State University\\
Ames, Iowa\\
}
\and
\IEEEauthorblockN{Girija Narlikar}
\IEEEauthorblockA{Bell Labs India\\
Alcatel-Lucent\\
}
\and
\IEEEauthorblockN{Gordon Wilfong}
\IEEEauthorblockA{Bell Labs USA\\
Alcatel-Lucent\\
}
}
\maketitle
}

\newpage

\begin{abstract}
Access networks, in particular, Digital Subscriber Line (DSL) equipment, are a significant source of energy consumption for wireline operators.   Replacing large monolithic DSLAMs with smaller remote DSLAM units closer to customers can reduce the energy consumption as well as increase the reach of the access network. This paper attempts to formalize the design and optimization of the ``last mile" wireline access network with energy as one of the costs to be minimized.  In particular, the placement of remote DSLAM units needs to be optimized.  We propose solutions for two scenarios.  For the scenario where an existing all-copper network from the central office to the customers is to be transformed into a fiber-copper network with remote DSLAM units, we present optimal polynomial-time solutions.  In the green-field scenario, both the access network layout and the placement of remote DSLAM units must be determined.  We show that this problem is NP-complete.  We present an optimal ILP formulation and also design an efficient heuristic-based approach to build a power-and-cost-optimized access network.  Our heuristic-based approach yields results that are very close to optimal. We show how the power consumption of the access network can be reduced by carefully laying the access network and introducing remote DSLAM units.

\end{abstract}

\section{Introduction}
\label{sec:intro}
Networks are traditionally designed and optimized for high performance.  However, optimizing their power consumption has become a secondary yet important goal.  Energy-related costs constitute a large part of the operating expenses for both wireless and wireline operators. These costs are especially severe in emerging economies, given a lack of access to reliable grid power, high ambient temperatures, and the high cost of maintaining power backups such as diesel generators.   For example, wireline access operators in India routinely run diesel generators on a daily basis; at some sites for several hours a day.

Wireline access networks are slowly evolving towards high-speed fiber laid all the way to the home as the ultimately scalable and energy-efficient architecture. However, given the extensive copper lines already in place, and the lower cost of copper-terminating customer premise equipment (CPEs), it may take several years before incumbent copper lines are replaced entirely with fiber.   Unfortunately, driving copper lines at high data rates and for long loop lengths consumes a significant amount of energy.    The line driver is one of the main sources of power consumption in DSL, and contributes close to 50\% of the power consumed by a DSLAM (`Digital Subscriber Line Access Multiplexer)~\cite{BLTJ}.

In current DSL access networks, an individual copper loop is placed between each customer and the nearest access node site, which houses the DSLAM. However, it takes a significant amount of power  to drive a copper line for each customer. The power level depends both on the user-requested bit-rate and the loop length of the DSL. Therefore, reducing the loop length will bring down the power consumption of the DSLAM equipment.  The DSL terminates at the  DSLAM on the network operator's side and these DSLAM units are responsible for driving the DSLs. DSLAMs are connected through optical fiber to the operator's data access network and fibers need a much lower level of power compared to copper wires.   Therefore, a straight-forward approach to save power is to place the DSLAM unit closer to customer premises and connect it to the nearest access node using optical fiber.  Thus, with shorter loop lengths, more customers can be served with higher data-rate services such as IPTV without destabilizing the lines or using excess transmit power.

Network equipment vendors around the globe have already started building remote DSLAM units which can be deployed at a suitable location near a group of customers.  These are currently used by operators to increase the reach of their access node to additional customers without compromising on data rates.  Because the remote DSLAM units don't need any real-estate (can be mounted on a roadside pole or in customer's basement) or active cooling (air conditioning), they have potential to bring down the operational expenses of a network operator significantly. However, the efficiency of this scheme depends on the strategic placement of remote DSLAMs in the service coverage area.

\begin{figure}[!h]
    \begin{center}
       \includegraphics[width=2.2in]{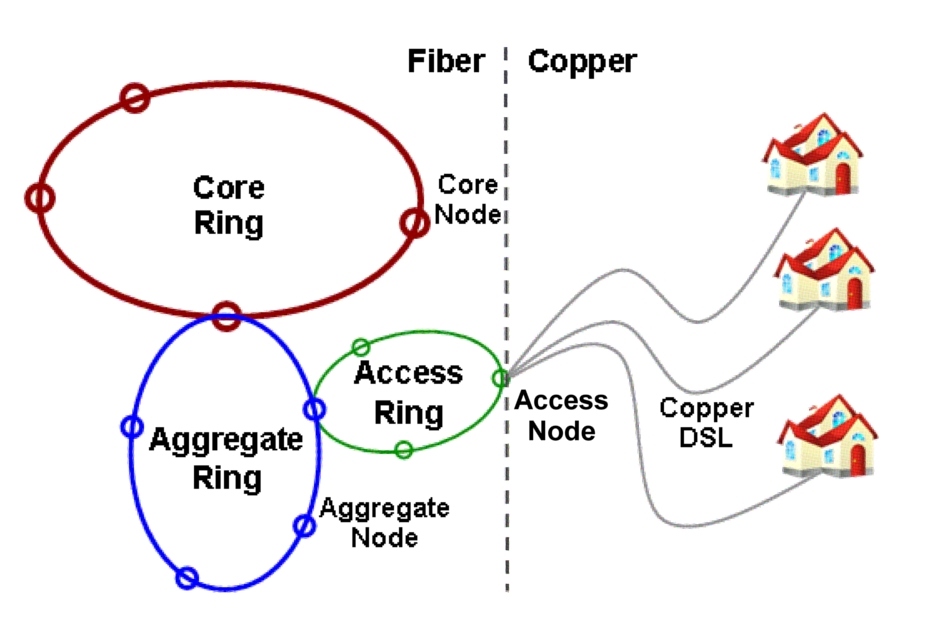}
        \caption{\bf Typical Metropolitan Access Network Structure}
        \label{fig:access_structure}
    \end{center}
\end{figure}
\begin{figure}[!h]
    \begin{center}
        \subfloat[]{\label{fig:dslam_place-a}
 \includegraphics[width=1.6in]{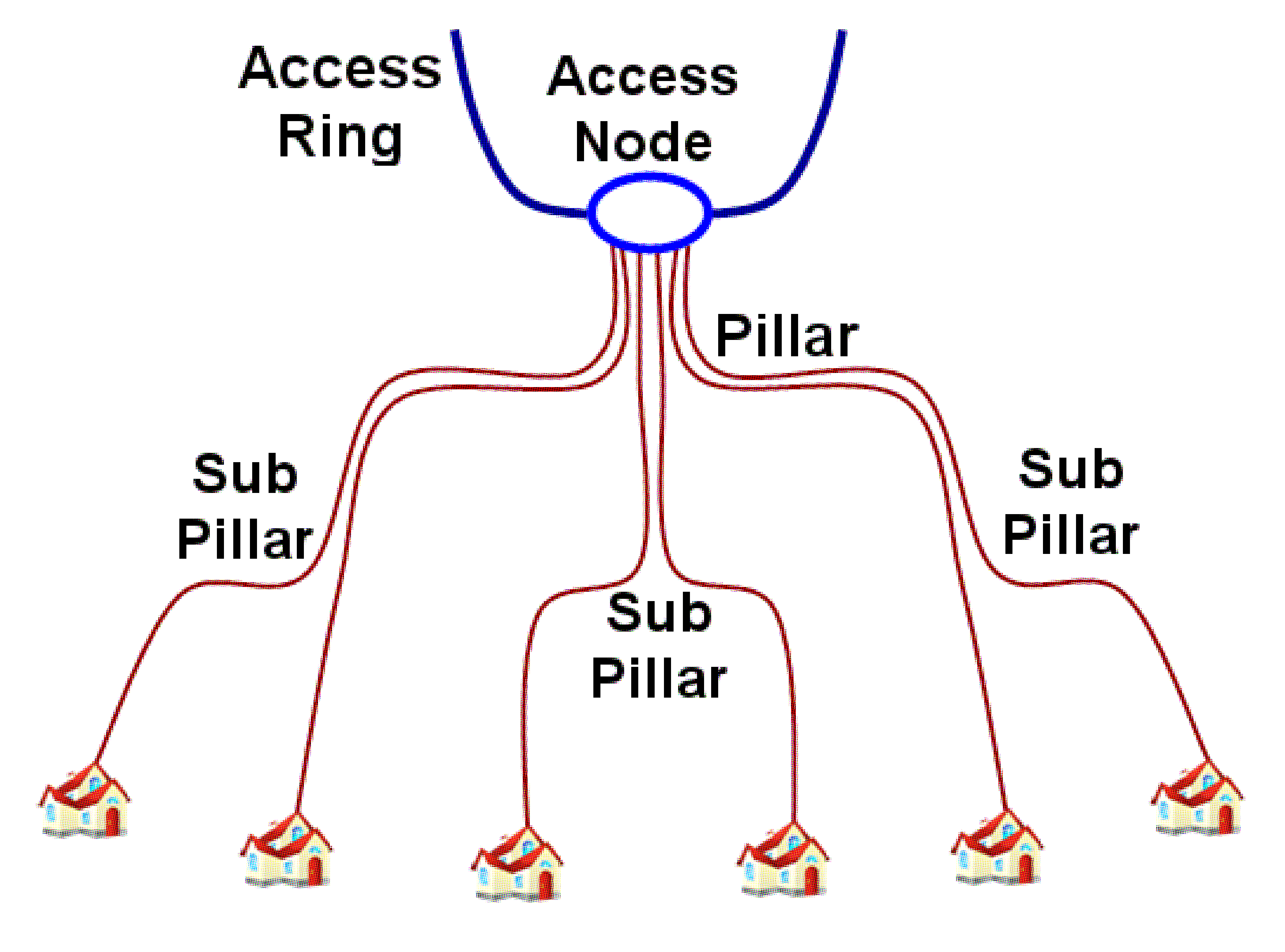}}
        \hspace{4.0in}
       \subfloat[]{
\includegraphics[width=1.6in]{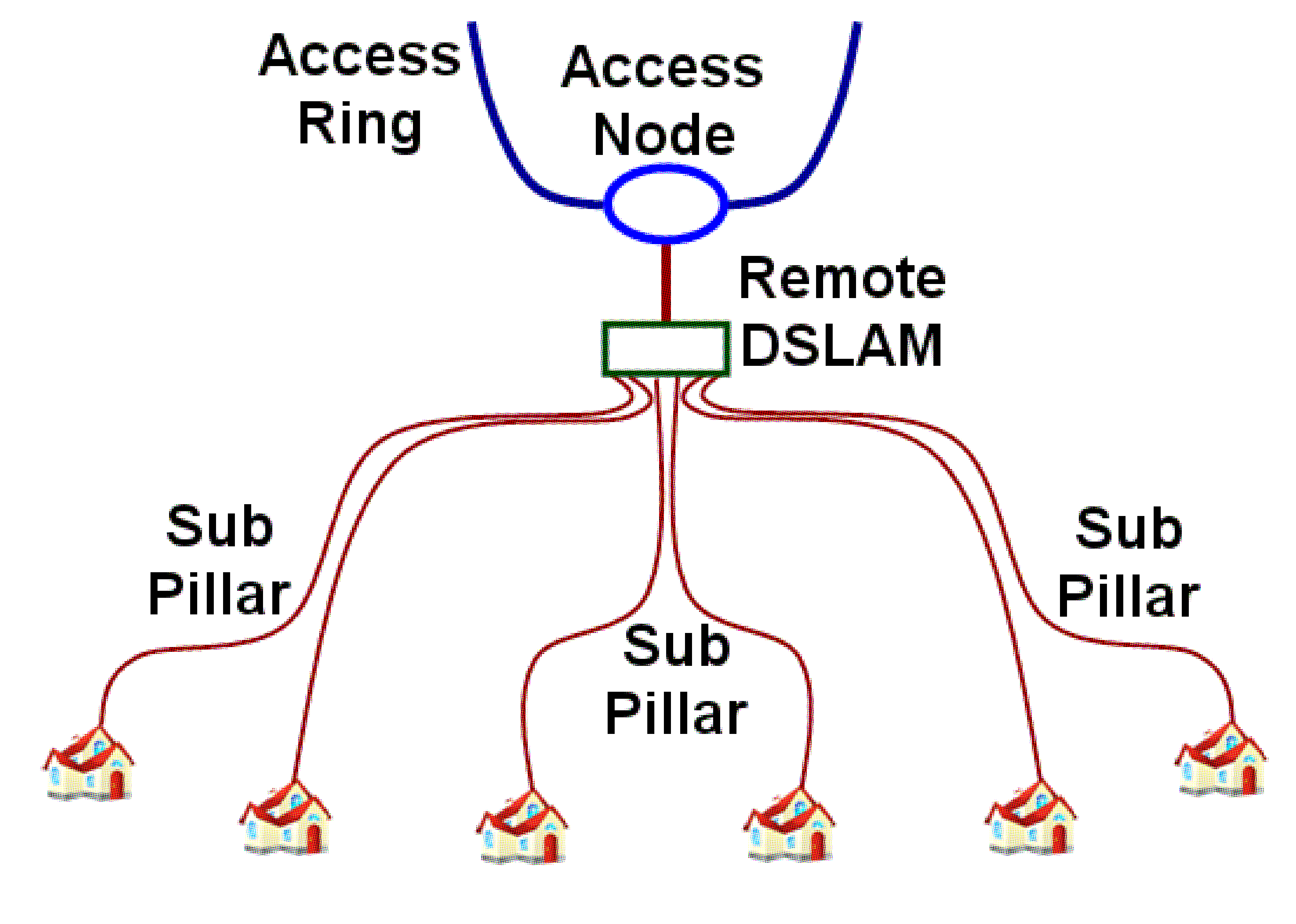}}
        \hspace{0.1in}
        \subfloat[]{
\includegraphics[width=1.6in]{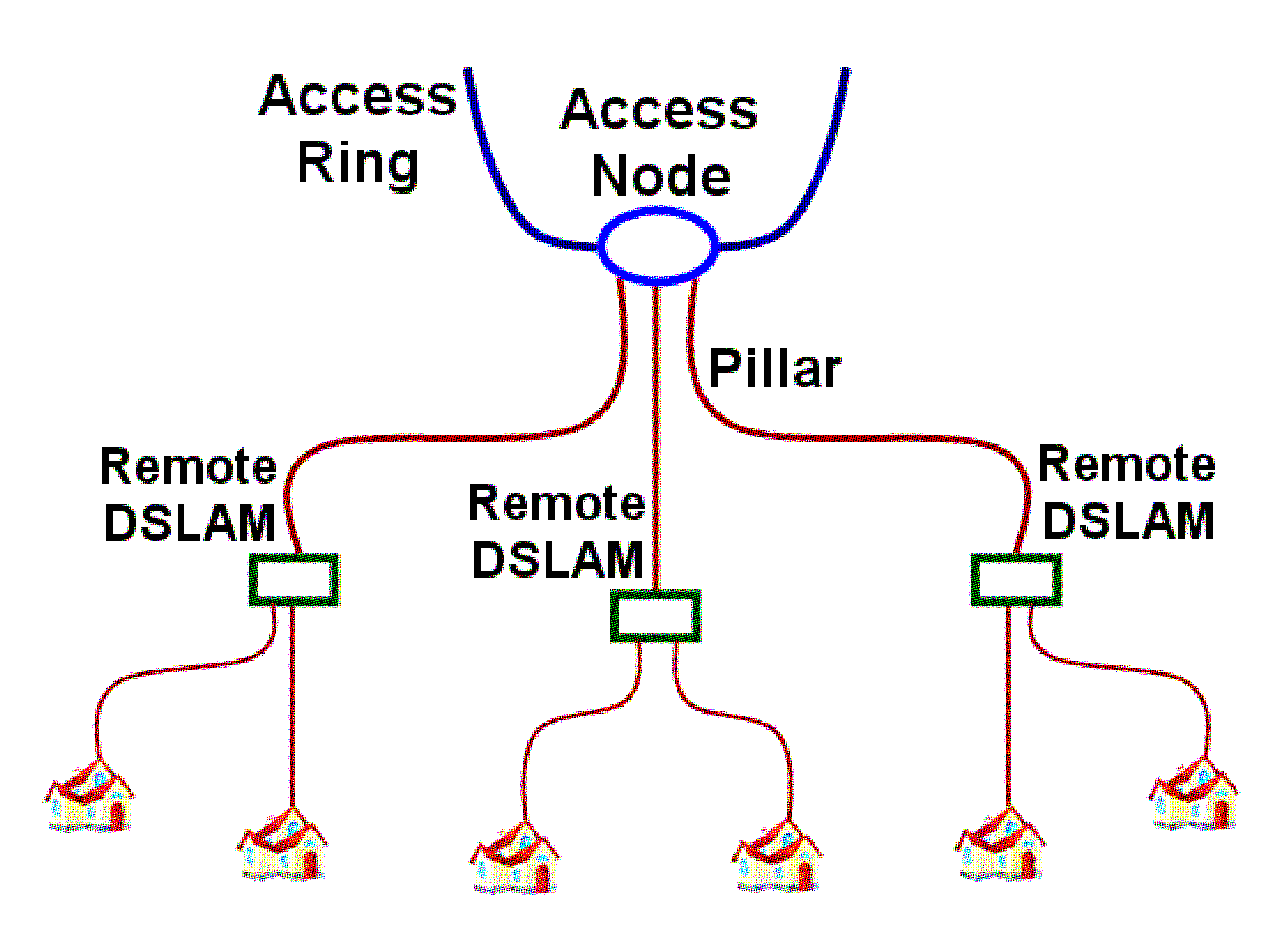}}
        \caption{\bf Remote DSLAM units in a Hierarchical Access Network}
        \label{fig:dslam_place}
    \end{center}
\end{figure}

This paper investigates the problem of designing and optimizing existing and green-field all-copper DSL networks by using remote DSLAM units.  Existing access networks contain large, monolithic DSLAMS that require a significant amount of energy to operate and cool.  We look at replacing these large DSLAMs with multiple strategically placed remote DSLAM units in the access network of an operator. We do not consider the entire access network as we do not intend to perturb the core of the network (e.g. Carrier Ethernet Network ring structure). Rather, we take into consideration the portion beyond an access node where copper lines are used to distribute the DSL connections.
In Figure \ref{fig:access_structure}, the area on the ``Copper'' side is the area of interest to us.
This part of the network is designed in a tree-based hierarchy, as shown in Figure~\ref{fig:dslam_place-a}.  For a large number of geographically scattered customers, there can be a choice of several alternative locations to place the remote DSLAM units (henceforth simply referred to as remote units).  Various local parameters such as the cost of  laying fiber, capex and opex of the remote units,  copper transmission, power costs, and customer locations can affect this choice.  Figure~\ref{fig:dslam_place} shows a few example remote unit placements.

There can be two different scenarios in which the placement of remote DSLAM can be carried out.  The first applies to an existing all-copper access network.  One must modify an existing access tree layout  and split the monolithic DSLAM unit into multiple smaller units distributed across the network.  This approach takes an all-copper tree and transforms it into a mix of copper and fiber, with the fiber starting from the root, that is, the access node. The fiber  terminates at some remote DSLAM unit, which initiates multiple copper loops to the end customers.  Typically, the additional cost of the remote units is justified by the increased reach of the access network to a much larger number of customers at high data rates.  We present an optimal polynomial-time algorithm to transform an all-copper tree into such a minimum-cost fiber/copper tree, using a dynamic programming solution.   We run this algorithm using network data based on a real operator's network. When the operator wishes to reach customers in an extended neighbourhood, placing remote units using our optimal algorithm results in significant improvements in both coverage (customers reachable) and power consumption.  For an example dataset, while an all-copper tree reaches only 50\% of the customers and consumes $~900mW$ of line driver power per customer, we can reach all the customers with the use of remote units  and reduce the power consumption to $~700mW$ per customer.

The second approach is to consider a green-field deployment.  We show that this problem is NP-hard, and present an ILP formulation.  While the ILP can solve very small problem sizes (~1 square kilometer in a semi-urban area), even slightly larger problem sizes become intractable. Therefore we also present a very fast and efficient heuristic based approach. This approach first builds an access tree (which determines where fiber or copper is to be laid), and then uses the previous algorithm to place remote DSLAM units as well as fiber and copper along this tree.  For datasets that can be solved using the ILP, our heuristic approach produces results very close to optimal.  For larger data sets, we show the benefit of using our algorithm on the customer coverage and power consumption.

\section{Related Work}

Traditionally DSL research has focused on maximizing data rates, increasing reach and providing line stability.  Therefore typically the transmit power is maximized on the DSLAM line cards.  Recently there have been demands from both the industry and governments to reduce broadband energy consumption~\cite{EUC}.  Therefore recent efforts have been made to reduce the energy consumption of DSLAM nodes by researchers~\cite{BL1,5199049,4737336,5199478,5208040}.  The ADSL2$+$ standard also now defines multiple low-power modes~\cite{adsl}. These efforts have been aimed at lowering the power consumed by individual copper loops.  Our approach is complimentary to them---given the physical parameters and costs in a given environment, we find a cost and energy efficient access architecture.

The placement of remote DSLAM units to create a hybrid fiber-copper network is somewhat similar
to the problem of designing hybrid wireless-optical access networks.  Here fiber is drawn as far as possible from the node site, and then having wireless access technologies to connect to the end user.  Optimal placement of wireless base stations is the equivalent problem in this scenario~\cite{1618569,490417}.  Here wireless propagation characteristics must be taking into account, especially in indoor environments~\cite{1374891}.  

Low-power networking research has typically focused on ad-hoc wireless systems with limited battery life. More recently, the energy-efficiency of networking infrastructure has been examined~\cite{863959,1559003}.  Underutilization of networking elements is a key contributor to wasted energy consumption~\cite{1559003, frontier}, and therefore scaling the power consumption of network elements based on network utilization is being pursued by researchers and networking vendors.  For example, as a network-wide optimization, turning off idle network elements and rerouting traffic can increase the energy efficiency at low traffic times~\cite{5426298,greener,4509688}.  Our approach of statically designing an energy-efficient access architecture is complimentary in spirit to these efforts.

\section{DSL Access Network Architecture}
\label{sec:network}
The statistics, network and customer data cited in this paper are based on the wireline access network of a large Indian operator.  As shown in Figure~\ref{fig:dslam_place}, the "last-mile" access network connects the access node site to the customers.  Traditionally individual copper loops run from the access node to each customer, and are bundled together in a hierarchical fashion.  The top-level bundles may contain a few hundred loops, while the lowest level close to the customer typically contains tens of loops.  Pillars and subpillars are used as junction points for the copper loops, to provide some flexibility and fault tolerance to the architecture.  These are passive equipment usually housed in a small shed.  At the leafs are distribution points located a few meters from individual customers' homes. Network operators always lay copper or fiber along the roads for ease of maintenance, repair and tracking.

A large node site, which also houses backhaul equipments and several routers and switches, may see only 10\% of its energy consumption coming from DSLAM's. In contrast, DSLAM's typically dominate the energy consumption of the equipments in a small access node site, contributing to as much as 80\% of it. These values are obtained from real time measurements carried out in a typical metropolitan area in India. The DSL line driver has been shown to contribute close to 50\% of the energy consumed by DSLAM nodes~\cite{BL1}.  Therefore reducing the loop lengths can significantly help reduce power consumption as well as support high data rates.

DSLAM equipment was traditionally installed as large monolithic nodes, with racks housing several line cards, supporting several hundred customers.  Such a centralized configuration implies that a significant number of customers are further from the node site.  Line driver power is typically set at a high level, so lines can be driven in a stable fashion.  Large numbers of such line cards packed in a small space results in significant cooling requirements.  Recently telecom vendors have been proposing a more distributed architecture, with smaller ``remote'' DSLAM units placed close to the customers.  This allows much shorter copper loop lengths, which can facilitate high data rates at low power.   Managing the larger number of remote units can be simplified by making them appear as remote line cards to the management layer. The downside, however, is that fiber needs to be laid (overhead or underground) up to the remote units.  We attempt to formalize this trade-off and develop algorithms to design the most cost-effective access network architecture.  In particular, we select the locations in the network suitable for placing remote units, and the customers who will be served by them.

\Paragraph{Terminology}  The following terminology will be used in the remainder of the paper.  For the tree redesign scenario, let $T$ be the existing tree layout. Let  $R$ denote the access node site, which is the root of the tree  $T$.  Let $V$ denote the set of nodes in $T$, and let $C \subset V$ be the set of customers in $T$.  In the greenfield tree layout problem the locality's roadways form a graph $G=(V,E)$; here each node $v\in V$ is a potential candidate to place a remote unit and each edge $(u,v) \in E$ is a segment of the road along which fiber or copper can be laid.   For simplicity we call all DSLAM units as remote units, including any units placed at $R$.   If a remote unit is laid at node $v$, then there must be a path from $R$ to $v$ along which fiber is laid.  There may also be additional copper loops along this path.

Let $power(u,v)$ be the cost of transmitting along a copper loop between nodes $u$ and $v$.  This cost of energy is a recurring operational expense. However, we translate it into a single upfront cost by computing  the total cost of units of energy (KWh) over a fixed period of time.  Let $rem(v)$ denote the cost of placing a remote unit at node $v \in V$, and let $D$ be its capacity.  Thus up to $D$ copper loops can be initiated from one remote unit.  Here $rem(v)$ includes the capital expenditure as well as rental/maintenance cost, again translated to a fixed upfront cost. Finally, let $fiber(v)$ denote the cost of laying fiber to the node $v$ from its parent. If we are installing one or remote units at $v$, the cost of laying fiber should also be added, otherwise this cost is $0$.

\section{Redesigning an existing copper-based tree layout}
\label{sec:tree}

In this section, we present a solution to the problem of redesigning an existing all-copper access tree into a tree with a mix of copper and fiber.  Note that the original and final trees may have multiple copper loops spanning the same edge: one copper loop is required to connect to each customer.  The final tree may have multiple copper loops as well as fibers along an edge.  One fiber is needed to connect each remote unit to the access node site. We attempt to minimize the overall cost of the tree. Our solution is a bottom-up dynamic programming based algorithm described below; the pseudocode is presented in Figure~\ref{fig:algo_code}.

In the original all-copper tree, for any subtree $T_s$ with $n$ customers in it, exactly $n$ copper loops enter the subtree along the edge connecting it to its parent. The tree redesign algorithm traverses the tree bottom up, and computes for each subtree the best way to reconfigure it, if some $c$ copper loops in the new tree enter the subtree along the edge connecting this subtree to its parent. This $c$ can be anywhere between $0$ and $n$, where $n$ is the number of customers in the subtree. The best configuration and its cost for each $c$ is therefore computed and stored in an array for each subtree (called $cost[]$), indexed by the number of incoming copper connections $c$.  If $c < n$, there must be one or more remote units placed within the subtree.  The best placement of these remote units for each value of $c$ is computed from the best configurations of its immediate subtrees. The final cost of the optimal transformed tree $T$ is $T\!\rightarrow\!cost[0]$, since only fiber enters the access node site.

Customers are assumed to own DSL-compliant modems. Therefore, customer nodes must have copper as the incoming medium. The tree starts off as all copper, so there is no additional cost of laying copper. However, there is a cost of laying fiber and placing remote units.  There is also a transmission power cost with the copper loops that remain in the transformed tree; the cost of this power for each copper loop is computed at and attributed to the node where the copper loop originates, that is, where their DSLAM is placed.  To compute the $cost[]$ array at a subtree rooted at a node $r$, the algorithm first computes a 2-d array $subtree\_cost[][]$. Here $subtree\_cost[k][c]$ is the optimal cost of distributing $c$ copper loops amongst the first $k$ subtrees of $r$.  This cost is computed using both the $cost[]$ arrays previously computed for each subtree and the row $subtree\_cost[k-1][]$ i.e. the row immediately before the current row being computed. Note that, if a subtree is a customer, we always force a copper to that line. For a node $r$ with $K$ subtrees, $subtree\_cost[K][c]$ therefore denotes the optimal cost of distributing $c$ copper loops among its children. To compute the value of $cost[c]$ at the node $r$ with $n$ customers under it, we compare the costs of distributing anywhere between $c$ and $n$ copper loops among its children, and picks the best configuration. When there are $c$ incoming copper loops at $r$, and it distributes $c' \geq c$ copper loops among its children,  it must house enough remote units at $r$ to drive the additional $c'-c$ copper loops.  Thus it incurs the cost of placing these remote units at $r$, and the power cost of driving the copper loops from these remote units at $r$.

When selecting the best way to distribute the $c'$ copper loops amongst customers under node $r$ with $c$ incoming copper loops,  we must make the following choice.   We need to decide which $c'-c$ customers under $r$ will be served by the remote unit(s) placed at $r$, and which $c$ customers will be served by one or more remote units above $r$ in the tree.  We do not yet know the locations of these remote units, and therefore cannot compute the exact power cost from the $c$ loops.   However, we do know that any copper loop originating from some node $r'$ above $r$ will need to traverse the path between $r'$ and $r$, in addition to traversing the path from $r$ to its customer.  Further, based on real data (for example, see Figure~\ref{fig:loop_power}), we know that the power cost $power(u,v)$ is a convex function for the loop length of $(u,v)$.  Therefore, if $e1, e2, e3, e4$ are edges in nondecreasing order of length, and $len(e1) + len(e4) = len(e2) + len(e3)$, then we know that $power(e1) + power(e4) \geq power(e2) + power(e3)$. We can therefore safely assign the $c$ shortest customer connections under $r$ to be served by remote units above $r$.

After this bottom-up traversal when we get the minimum cost $T\!\rightarrow\!cost[0]$, the tree is again traversed in a top-down manner and this time the configuration for each node is set according to the best configuration chosen at their parent. For example, we start by choosing the configuration corresponding to $cost[0]$ at root, because this is the only possible configuration for access node site. Suppose, the root has $K$ children and the chosen configuration distributes respectively $c_1, c_2 \ldots c_K$ copper lines to these nodes. We then select the configuration of $cost[c_1]$ at first child, configuration of $cost[c_2]$ at second child and likewise $cost[c_K]$ at the $K^{th}$ child. After this, we continue to set the configurations of nodes further below the tree until all nodes are visited.

\begin{figure}[t]
\begin{center}
\begin{minipage}[t]{5in}
\begin{tabbing}
xx\=xx\=x\=x\=x\=xx\=xxxxxxxxxx\= \kill
Traverse $T$ in depth-first-order, and run\\
the following code for each subtree $T_s$: \\
\> $r := root(T_s);$ \\
\> $n := customer\_count(T_s)$;\\
\>$K := subtree\_count(T_s)$;\\
\> $subtree\_assignment(T_s, n);$\\
\> for $c:=0\; to\; n$: \\
\> \>  $T_s\rightarrow cost[c] := $\\
\>\>\>\>$min_{j := c\; to \;n}\bigl( subtree\_cost[K][j] + fiber(r) +$\\
\>\>\>\>\> $rem(r) * \lceil(j-c)/D\rceil +\sum_{v\in end(j-c)} power(r,v)\bigr)$ \\
\\
Traverse $T$ in post-order and run\\
the following code for each node $r$: \\
\> if $r$ is root of the tree\\
\> \> $r \rightarrow configuration := 0$;\\
\> else\\
\> \> $p := parent(r) \rightarrow configuration$;\\
\> \> $c := copper\ assigned\ to\ r\ in\ p$;\\
\> \> $r \rightarrow configuration := c$;\\
\\
subroutine $subtree\_assignment(T, n):$\\
\>Let  $T_{1}, T_{2},\ldots,T_{K}$ be the K subtrees of $T;$\\
\>for $c:=0\; to\; n$: \\
\>\>$subtree\_cost[1][c] := T_{1}\!\rightarrow\! cost[c];$\\
\>for $c:=0\; to\; n$: \\
\>\> for  $k := 2\;to\;K$:\\
\>\>\>$subtree\_cost[k][c] = min_{j:=0\;to\;c}\bigl( T_k\!\rightarrow\! cost[j] + $\\
\>\>\>\>$subtree\_cost[k-1][c-j]\bigr);$\\
\\
\end{tabbing}
\end{minipage}
\end{center}
\vspace{-0.2in}
\caption{Pseudocode for the algorithm to transform a copper tree into a fiber-copper tree by adding remote units.
}
\label{fig:algo_code}
\end{figure}

\Paragraph{Complexity} Consider an original copper tree with $n$ nodes, where each node has at most $K$ children. Our algorithm takes time $O(n^2\cdot K)$ work and $O(n\cdot K)$ space to compute the optimal configuration at each node in terms of the array $subtree\_cost[][]$. It also maintains a persistent $O(n)$ cost array $cost[]$ for every node. The algorithm therefore requires a total of $O(n^3\cdot K)$ time and $O(n^2)$ space.

\Paragraph{Correctness} The output of the algorithm is a valid access network tree due to the following constraints maintained by the algorithm - 1) Customer nodes are supplied with copper line hence no customer will end up receiving fiber, 2) A required number of remote units are placed at each node, therefore all copper lines have a serving DSLAM, and 3) All remote units are connected to access node site by placing fiber between them.

\Paragraph{Optimality} We present an informal argument to show that the algorithm always finds the best configuration for an access tree. Suppose this is not true and the output is a sub-optimal configuration. This sub-optimal configuration can be improved in two ways - 1) By choosing a different number of incoming copper lines for a subtree - but this will increase the total cost because we are already choosing copper lines at each subtree in order to minimize the total cost at root. And 2) By assigning the copper lines to a different set of customers - but this will also increase the total cost because the power function is a convex function and we are already choosing the shorter loops to be served by the coppers originating higher in the tree. Hence the configuration can not be improved or it is indeed the optimal configuration, contradicting our initial assumption.

\Paragraph{Budget Restricted Optimization} The algorithm described above optimizes between the expenses of new deployment and the saving of transmit power over copper lines. Instead, consider a  network operator who is willing to spend a certain amount to optimize the network and wants to know what is the best configuration possible with that expense. The above algorithm can be modified to achieve this optimization, but in this case instead of finding the best assignment of copper lines, we find the best assignment of budget among nodes. In this case, we also specify a budget amount as an input to the algorithm.

\subsection{Experiments}
\label{sec:tree-expts}

\begin{figure}[!h]
    \begin{center}
       \includegraphics[width=1.6in]{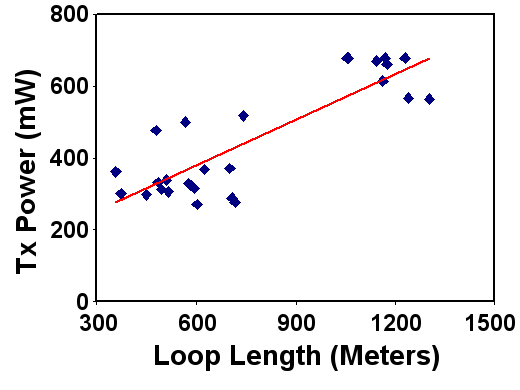}
 \end{center}
        \vspace*{-0.1in}
        \caption{Line driver power vs loop length assumed in our experiments, assuming a 25db SNR gap.  These are results of simulations of a typical urban bundling scenario.}
    \label{fig:loop_power}
\end{figure}

We now present experiments based on a real urban all-copper access tree; the network layout is derived from an existing access network in a large Indian city, while the costs are based on realistic values of today (see Table~\ref{tab:tree-parameter}).  We had data for only 150 sample customers; a real access node site typically serves several hundred customers.  We will scale up this number for some of the experiments later in this section.   The remote unit cost is estimated as  the {\em additional} per-unit incremental cost of replacing an existing DSL line card from a monolithic DSLAM with a remote unit.  The remote unit carries out the same functionality as a line card, but includes additional overheads of providing power and casing.  The cost of digging fiber tends to be higher than copper (deeper trenches), but the material cost of copper is higher.  Hence the total cost for laying both are comparable in our scenario. Figure~\ref{fig:loop_power} shows a typical relation between line driver power and loop length for a fixed data rate of 12Mbps and an SNR gap of 25db.  Higher SNR gaps imply better line stability but also higher line driver power.  In this scenario, lower gaps result in a fairly flat and low power curve, while higher gaps result in constant high power consumption.  For lower gaps  (under 20db), sustaining this data rate is possible only for very short loops.  Above a loop length of 1500m, we assume that sustaining 12Mbps with 25db SNR gap level is not possible.  Therefore all loop lengths must be up to 1500m. For lines longer than this limit, remote DSLAM units are necessary to shorten the copper loop lengths.

 \begin{table}
 \centering
 \begin{tabular}{|l || c|}
  \hline \textbf{Parameters} & \textbf{Value}\\
  \hline Maximum loop length & 1500 m\\
  \hline Cost of laying fiber (option 1) & \$6 per meter\\
\hline Cost of laying fiber (option 2) & \$0.50 per meter\\
\hline Cost of remote DSLAM unit & \$2000\\
\hline Capacity (\#copper ports) of  remote DSLAM unit & 50\\
\hline Cost of energy & \$0.20 per kWh\\
\hline Time period to recoup capital expense & 3 years\\
   \hline
\multicolumn{2}{|c|}{Customer network parameters}\\
\hline Number of customers & 150\\
\hline Dimensions of copper tree layout & 1km x 1km\\
\hline Type of Scenario & dense urban \\
\hline Per user data rate & 12 Mbps\\
\hline
   \end{tabular}
   \caption{Parameter values for tree reconfiguration}
   \label{tab:tree-parameter}
\end{table}

\begin{figure}[!h]
    \begin{center}
        \includegraphics[width=1.6in]{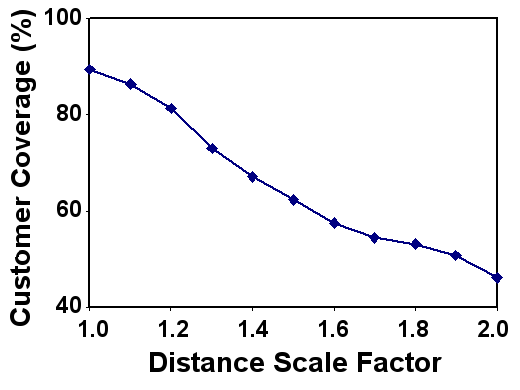}
    \end{center}
    \vspace*{-0.1in}
    \caption{Percentage of customers reachable by a copper loop from the access node site without remote units with the increasing distance.}
    \label{fig:scale_cover}
\end{figure}

\begin{figure}[!h]
    \begin{center}
        \subfloat[]{
            \includegraphics[width=1.6in]{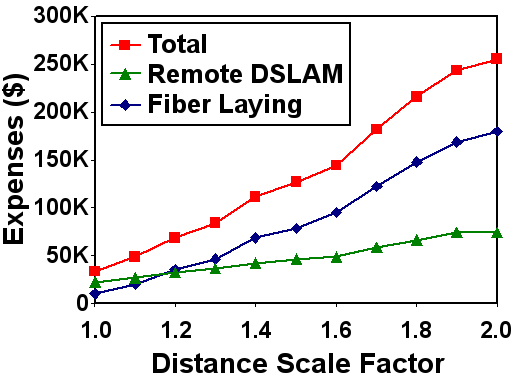}
            \label{fig:distance_expense}}
        \hspace{0.05in}
        \subfloat[]{
            \includegraphics[width=1.6in]{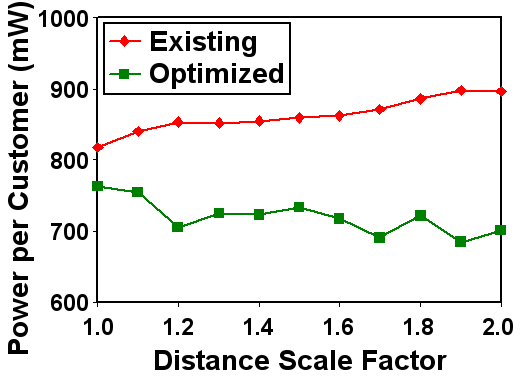}
            \label{fig:distance_power}}
    \end{center}
    \vspace*{-0.1in}
    \caption{Effect of scaling network edge lengths on the costs and power consumption of the optimal access tree.}
    \label{fig:distance_plot}
\end{figure}

We now present our tree redesign algorithm on the existing all-copper tree. Most loop lengths in the existing access tree are fairly small , as service is not available to customers further from the node site. Since very few customers are at a distance of over 1500m, we decided to scale up the distances of each edge in the original access tree.  Figure~\ref{fig:scale_cover} shows the percentage of customers who can be reached by an all-copper tree with no remote units, as the scale factor is increased. Thus, less than 50\% of the customers can be reached from the access node site with copper loops when distances are scaled up by a factor of 2; the rest are too far away  and need to connect to a remote DSLAM unit.
Figure~\ref{fig:distance_plot} shows the effect of our algorithm on the all-copper tree. Figure~\ref{fig:distance_expense} shows the increasing expenses of network in terms of remote DSLAM unit, fiber laying and total costs. Figure~\ref{fig:distance_power} shows the saving in line driver power achieved by the optimization---at higher distances as much as $200$ milliwatts (over $20\%$) power per line can be saved.

\begin{figure}[!h]
    \begin{center}
        \subfloat[]{
            \includegraphics[width=1.6in]{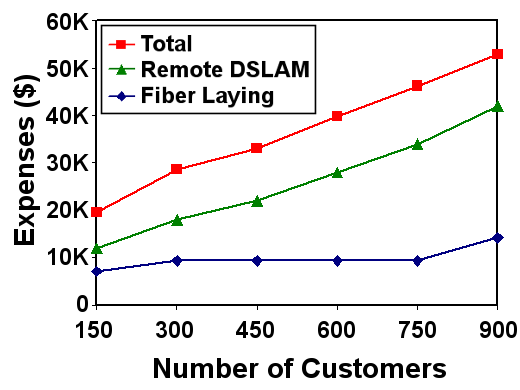}
            \label{fig:user_expense}}
        \hspace{0.05in}
        \subfloat[]{
            \includegraphics[width=1.6in]{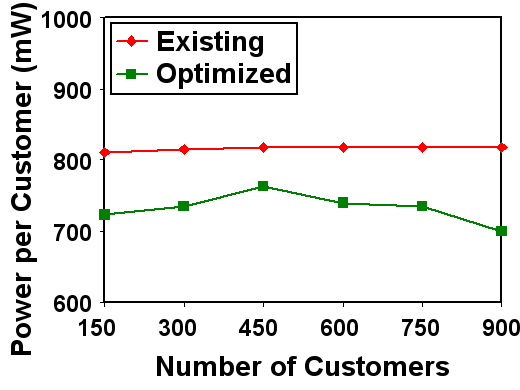}
            \label{fig:user_power}}
    \end{center}
    \vspace*{-0.1in}
    \caption{Effect of increasing network subscribers on the costs and power consumption of the optimal access tree.}
    \label{fig:user_plot}
\end{figure}

We now show the effect of varying the number of customers in the access network. We increase the number of customers in the network from $150$ to $900$ and execute the tree redesign algorithm on it. The results are shown in Figure~\ref{fig:user_plot}. Figure~\ref{fig:user_expense} shows the increasing expenses required to achieve an optimal configuration with different number of users and Figure~\ref{fig:user_power} shows the line driver power saving achieved by the optimal configuration. Note that, in this case the distances are not scaled up and more than $90\%$ of the customers are within the reach of copper loops.  Our algorithm still achieves around $100$ milliwatts saving per line.

\begin{figure}[!h]
    \begin{center}
        \subfloat[]{
            \includegraphics[width=1.6in]{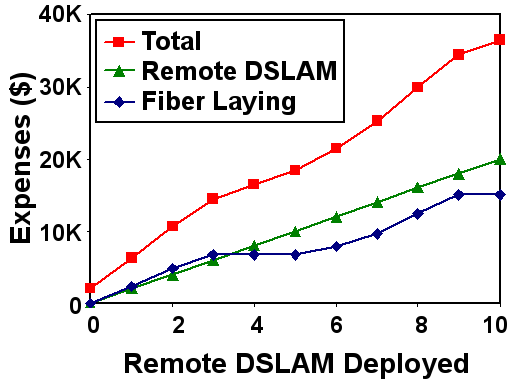}
            \label{fig:budget_expense}}
        \hspace{0.05in}
        \subfloat[]{
            \includegraphics[width=1.6in]{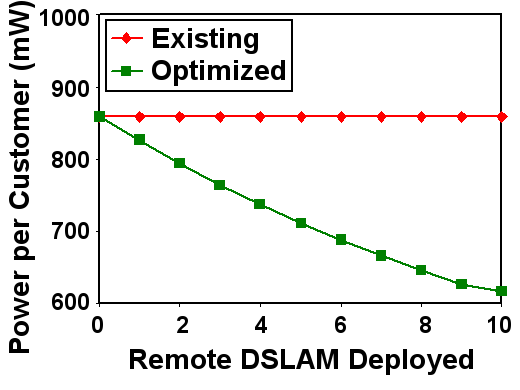}
            \label{fig:budget_power}}
    \end{center}
    \vspace*{-0.1in}
    \caption{Effect of increasing remote units on the costs and power consumption of the optimal access tree.}
    \label{fig:budget_plot}
\end{figure}

Next, we present the results from the budget restricted tree redesign algorithm.  For simplicity, we mention the number of remote units that we want to deploy in the network, the combined budget of remote unit and fiber laying can be specified similarly. Figure~\ref{fig:budget_expense} shows the expenses incurred by the optimization with increasing remote DSLAM unit deployment. Figure~\ref{fig:budget_power} shows the saving in line driver power achieved with these remote DSLAM units. As expected, the saving increases with the number of remote DSLAM units and we achieve a saving of close to $300$ milliwatts per line with $10$ remote units.

\begin{figure}[!h]
    \begin{center}
        \includegraphics[width=1.6in]{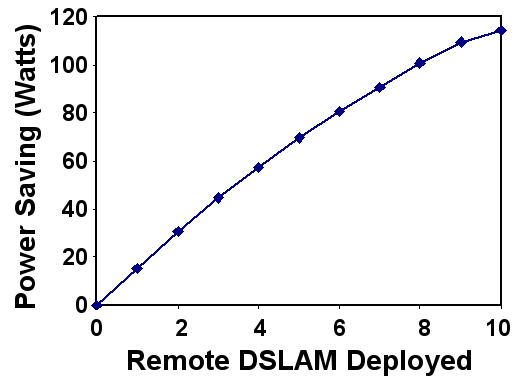}
    \end{center}
    \vspace*{-0.1in}
    \caption{Total cost saving obtained with increasing budget.}
    \label{fig:budget_saving}
\end{figure}

\comment{
Finally, Figure~\ref{fig:power_remcost} shows how the remote unit cost affects the power consumption and total network cost of the optimally reconfigured tree.

  \begin{figure}[!h]
    \begin{center}
       \includegraphics[width=2.5in]{RemCostVary.png}
         \end{center}
    \vspace*{-0.1in}
    \caption{Variation of power and number of remote units (REMs) in the access network as determined by our optimal algorithm, as the remote unit cost is varied.  When remote units are very inexpensive, many are used and power consumption is low. As the remote units become expensive, fewer get utilized and power consumption rises. }
   \label{fig:power_remcost}
\end{figure}

 \begin{figure}[!h]
    \begin{center}
       \includegraphics[width=2.5in]{RemCost_TotalCost.png}
         \end{center}
  \vspace*{-0.1in}
  \caption{Variation of total network cost as a function of the cost  of remote units (REMs) in the access network.  Here the network cost is computed by our optimal algorithm.  When remote units are very inexpensive, they dominate the total network cost.  }
     \label{fig:cost_remcost}
\end{figure}
}

Finally, Figure~\ref{fig:budget_saving}  shows the total saving achieved by the budget restricted optimization algorithm.  Note that the power savings from the use of remote DSLAM units alone cannot fully justify their additional capital expense. However, remote units can also enable high bit-rate DSL services for a much larger set of customers than is feasible with pure copper access trees.
To reach  customers further away from the node site, the operator must set up a whole new access node site. The cheaper and simpler alternative is to introduce remote DSLAM units within the network.  This task is typically carried out manually; our  algorithm automates this process of remote DSLAM unit placement in an optimal cost and power-efficient manner.

\section{Designing a green-field access network}

We now describe the problem of designing an access network with no existing copper laid out---both the fiber and copper portions of the tree need to be designed. The growing broadband penetration in several emerging economies often leads to such a scenario.  We assume here that the location of the central office and the customers are given as input.  Further,  we are given the segment-wise street map which determines routes along which fiber or copper may be laid.  The cost of laying fiber or copper may vary widely from road segment to road segment.  Further, we assume that some of the segment endpoints are candidates for placing remote DSLAM units.  To assure high quality and bit rates, we assume a hard limit on the length of the copper loops---any customers that are too far from the node site are required to connect to it through a remote DSLAM unit. The goal of this  section is to design the least-cost fiber-copper access network that connects all the customers to the node site.

\subsection{Problem Formulation}

In the green-field version, we are given a graph $G=(V,E)$
where the edges represent road sections along which a trench can
be dug, there is a designated node representing the central office,
and the remaining nodes represent either road intersections or customers.
There are costs along each edge giving the copper transmission power on
on that edge, the cost to dig a trench along
that stretch of road, the cost to lay copper within such a trench
and the cost to lay fiber within the trench.
Some nodes are designated as potential locations for a remote DSLAM and
these nodes have a cost indicating the cost to install a remote
DSLAM at that node.
DSLAMs have a limit on the number of customers that they can service.
Our goal is to determine a subset of the nodes at which to place
remote DSLAMs, a  tree of edges along which trenches will
be dug to allow for fiber from the central office to each remote DSLAM
and for each customer, copper from a remote DSLAM to the customer.

If the costs for laying fiber and copper are 0 and the cost to install
a remote DSLAM is large (essentially infinite)
then the cost of such a design is just the
cost to dig a tree of trenches to connect the customers with the central office.
Thus a special case of our problem is the Steiner tree problem, a well-known
NP-complete problem~\cite{GareyJ79}.
Therefore the green-field design problem is NP-complete as well.

Since the green-field design problem is NP-complete, we develop
an integer linear program (ILP) and a heuristic approach.
We will use the optimal results produced by the ILP to measure the quality
of the heuristic solutions for smaller, more feasible datasets.

\subsection{Integer Linear Program}

 \begin{table}
\centering
 \begin{tabular}{|c || l|}
  \hline \textbf{Parameters} & \textbf{Description}\\
  \hline $E$ & The set of edges in the graph.\\
  \hline $U$ & The set of customers, where $U \subseteq V$.\\
  \hline $R$ & Set of DSLAM candidate locations, $R \subseteq V$, $R \cap U = \phi$\\
  \hline $S$ & Central office node, $S \cap R =\phi$ \\
  \hline $R_N$ & The set of non-candidate points, $R_N$ = $V \setminus(S \cup U \cup R)$\\
   \hline $Dig(e_{xy} )$ & The digging cost of edge $e_{xy} \in E$.\\
   \hline $f_t^c(x,y)$ & Copper transmission power from $x$ to $y$.\\
   \hline $f_I^c(e_{xy})$ & Copper installation cost on $e_{xy}$.\\
   \hline $f_I^o(e_{xy})$ & Fiber installation cost on $e_{xy}$.\\
   \hline $N(x)$ & The set of neighbors of $x$.\\
   \hline $D$/$C_D$ & DSLAM capacity / cost.\\
   \hline $L_{xy}$ & length of $e_{xy}$.\\
   \hline $L$ & maximum distance of copper from DSLAM to customer.\\

   \hline
   \hline \textbf{Variables} & \textbf{Description}\\
   \hline $c_{xy}[i]$ & Indicator whether copper to customer $i$ installed from $x$ to $y$.\\
   \hline $n_{xy}^o$ & The number of fiber installed from node $x$ to node $y$.\\
   \hline \multirow{3}{*}{$T_{xy}$} & Indicator whether a trench is dug on $e_{xy}$.\\
                  & \begin{numcases}{T_{xy}  =}
                    1 & if a trench is dug \nonumber\\
                    0 & Otherwise\nonumber
                  \end{numcases}\\
                   & \\
   \hline $d_{i}$ & Number of DSLAMs installed at node $i\in R.$\\
   \hline
   \end{tabular}
   \caption{ILP Parameters and Variables}
   \label{tab:parameter_variable}
\end{table}


The parameters and variables of the ILP are shown in Table~\ref{tab:parameter_variable}.
The ILP's objective is to minimize:
  \begin{eqnarray}
        &&\underbrace{\sum_{ e_{xy} \in E} T_{xy}\cdot Dig(e_{xy} )+\sum_{i\in R} d_i \cdot C_{D}}_{I}+\underbrace{\sum_{e_{xy}\in E} n_{xy}^{o}\cdot f_I^o(e_{xy}}_{II}) \nonumber\\
&&+\underbrace{\sum_{i\in U} \sum_{ e_{xy} \in E} c_{xy}[i]\cdot \{f_{t}^c(x,y) + f_I^c(e_{xy})\}}_{III}\nonumber
  \end{eqnarray}
$where:$\\
\hspace*{1cm}\emph{I. Digging and DSLAM installation cost.}\\
\hspace*{1cm}\emph{II. Fiber installation cost.}\\
\hspace*{1cm}\emph{III. Copper Tx power and installation cost.}\\
subject to the following constraints:
  \begin{itemize}
    \item $\forall x \in R_N$, the number of incoming copper/fiber lines should equal the number of outgoing copper/fiber lines:
        \[
            \sum_{y\in N(x)}c_{xy}[i] - \sum_{y\in N(x)}c_{yx}[i] = 0  \hspace*{.25cm} \forall x \in R_N, \forall i\in U
        \]
        \[
             \sum_{y\in N(x)}n_{xy}^o - \sum_{y\in N(x)}n_{yx}^o = 0  \hspace*{1cm} \forall x \in R _N
        \]
    \item $\forall i\in R$, if $d_i$ DSLAMs are installed, then the number of incoming fibers is $d_i$ more than the number of outgoing fibers.  Also the number of net outgoing copper lines can be at most the capacity that $d_i$ DSLAMs can support.


        \[
            \sum_{j\in N(i)}n_{ji}^o - \sum_{j\in N(i)}n_{ij}^o = d_i \hspace*{2cm} \forall i\in R
        \]
        \[
            \sum_{k \in U} \Bigg(\sum_{j\in N(i)}c_{ij}[k] - \sum_{j\in N(i)}c_{ji}[k]\Bigg)\leq d_i \cdot D  \hspace*{0.5cm} \forall i \in R
        \]
    \item Since $S$ is not a candidate location, no copper lines can go in and out of $S$ and no fiber go into $S$ as well.  Additionally, the number of fiber coming out of $S$ should be the total number of installed DSLAMs.
        \[
        \label{equ:central 1}
            \sum_{i \in U} \sum_{y\in N(S)}c_{Sy}[i] =\sum_{i \in U} \sum_{y\in N(S)}c_{yS}[i]=  \sum_{y\in N(S)}n_{Sy}^o= 0
        \]
        \[
        \label{equ:central 2}
            \sum_{y\in N(S)}n_{Sy}^o = \sum_{i\in U} d_i\hspace*{2cm}
        \]
    \item Each customer should consume only 1 copper line.  The number of incoming fiber should equal to the number of outgoing fiber.
        \[
            \sum_{y\in N(i)}c_{yi}[i] = 1,  \sum_{y\in N(i)}c_{iy}[i] = 0 \hspace*{1cm}  \forall i \in U
        \]
        \[
            \sum_{y\in N(i)}c_{yi}[j] - \sum_{y\in N(i)}c_{iy}[j]= 0 \hspace*{1cm} \forall i, j \in U, j\not= i
        \]
        \[
            \sum_{y\in N(i)}n_{yi}^o - \sum_{y\in N(i)}n_{iy}^0= 0 \hspace*{1cm} \forall i, j \in U, j\not= i
        \]
    \item Copper and fiber can only be installed if a trench has been dug.  The allowable length of copper lines from a DSLAM to a customer is bounded. \\
    $\forall e_{xy}\in E, i \in U:$
    \[
        c_{xy}[i]+ c_{yx}[i] + n_{xy}^o+ n_{yx}^o \leq T_{xy}\cdot(|R|+|U|)
    \]
        \[
             \sum_{e_{xy} \in E} c_{xy}[i] \cdot L_{xy} \le L \hspace*{2cm}
        \]

  \end{itemize}

Note that we have assumed the transmission power $f_t^c(x,y)$ over copper to be linear and additive, that is, the total power consumed over a series of segments is the sum of transmission powers over each segment.  In reality, how a copper loop is bundled with other copper loops determines the crosstalk effects, and will therefore influence the transmit power to some degree.  Further, while Figure~\ref{fig:loop_power} indicates an acceptable approximation using a linear curve, it does not pass through zero.  Hence when a long copper loop is laid over a series of many small segments, the actual power consumed will be less than the power obtained by adding up power over the individual segments.

The ILP turns out to be intractable over any reasonably large inputs, and therefore in the next section we present a simple and fast heuristic to solve the greenfield problem. Our above approximate segment-wise linear power model is also adopted by the heuristic approach only to make a fair comparison against the ILP.  For a real deployment it can use any arbitrary and more precise power model.

\subsection{Heuristic Approach}

We now describe a heuristic approach to solve the green-field design problem.

Typically the cost to dig a trench will be a major factor in the overall cost
of a design.  The higher the trenching cost, the closer the overall problem resembles the minimum-cost Steiner tree problem.
Therefore a reasonable greedy approach would be to first find a low cost (in
terms of the trenching costs) tree $T$, that is, a Steiner tree
that connects all the customers to the central office.
Therefore our heuristic approach first constructs a low-cost Steiner tree.  Given such a Steiner tree $T$, we then apply the dynamic programming
algorithm described in Section~\ref{sec:tree} to obtain a complete design.
In the dynamic programming step, we assign a zero cost to the trenching, since that is accounted for in the Steiner tree cost.  Instead, the dynamic program balances the cost of the remote units against the energy saved.

Multiple Steiner tree approximation algorithms exist.  We use a simple algorithm with a 2-approximation ratio~\cite{algos}.  It builds a complete graph of customers and the central office based on pairwise shortest paths , then constructs a minimum spanning tree (MST), and maps it back to the original graph.  We will refer to this as the ``Steiner" method.  We also use an alternative method of building an MST and then deleting subtrees that do not contain any customers.  Henceforth this approach will be labeled as ``MST".  We also plan to implement the more complex 1.55-approximation Steiner tree algorithm by Robins and Zelikovsky~\cite{338638}
for the full paper.

There may always be optimal tree layouts that the ILP discovers but are not found by the heuristic approach. For example, the ILP may decide to place a remote unit at a node, then run some of the copper loops from the node {\em up} the tree and down to another subtree.  As shown in the next section, the heuristic results are nevertheless near-optimal.

\comment{}
\section{Experiments}
\label{sec:exp}

We tested out the ILP and heuristic algorithms on a datasets generated for two Indian cities---Bhopal and Kolkata.  We extracted a square area from the street map of each city, and used the street segments as edges; their end points are candidate locations for placing remote DSLAMs.  The customer locations were chosen at random from the segment ends.  Table~\ref{tab:graph_descriptors} lists the various datasets and their details.

\begin{table}[!h]
    \centering
    \begin{tabular}{|c|c||c|c|c|c|}\hline
        \textbf{City} & \textbf{Data Set} & \textbf{Dimen-} & \textbf{No. of} & \textbf{No. of} & \textbf{No. of} \\
        \textbf{Name} & \textbf{Name} & \textbf{sion (km)} & \textbf{Nodes} & \textbf{Edges} & \textbf{Customers} \\ \hline

        \multirow{5}{*}{Bhopal} & B0.5 & 0.5 & 20 & 38 & 15 \\ \cline{2-5}
                                & B0.8 & 0.8 & 53 & 108 & 30 \\ \cline{2-5}
                                & B1.0 & 1.0 & 91 & 188 & 50 \\ \cline{2-5}
                                & B2.0 & 2.0 & 583 & 1266 & 300 \\ \cline{2-5}
                                & B3.0 & 3.0 & 1026 & 2242 & 600 \\ \cline{2-5}
                                & B4.0 & 4.0 & 1677 & 3692 & 800 \\ \hline
        \multirow{5}{*}{Kolkata}& K0.5 & 0.5 & 20 & 38 & 12 \\ \cline{2-5}
                                & K0.8 & 0.8 & 46 & 92 & 32 \\ \cline{2-5}
                                & K1.0 & 1.0 & 92 & 184 & 50 \\ \cline{2-5}
                                & K2.0 & 2.0 & 667 & 1462 & 300 \\ \cline{2-5}
                                & K3.0 & 3.0 & 1522 & 3398 & 600 \\ \cline{2-5}
                                & K4.0 & 4.0 & 2317 & 5190 & 800 \\ \hline
    \end{tabular}
    \caption{Details of graphs used in our experiments.}
    \label{tab:graph_descriptors}
\end{table}

\begin{figure}[!h]
    \begin{center}
        \subfloat[]{
            \includegraphics[width=2in]{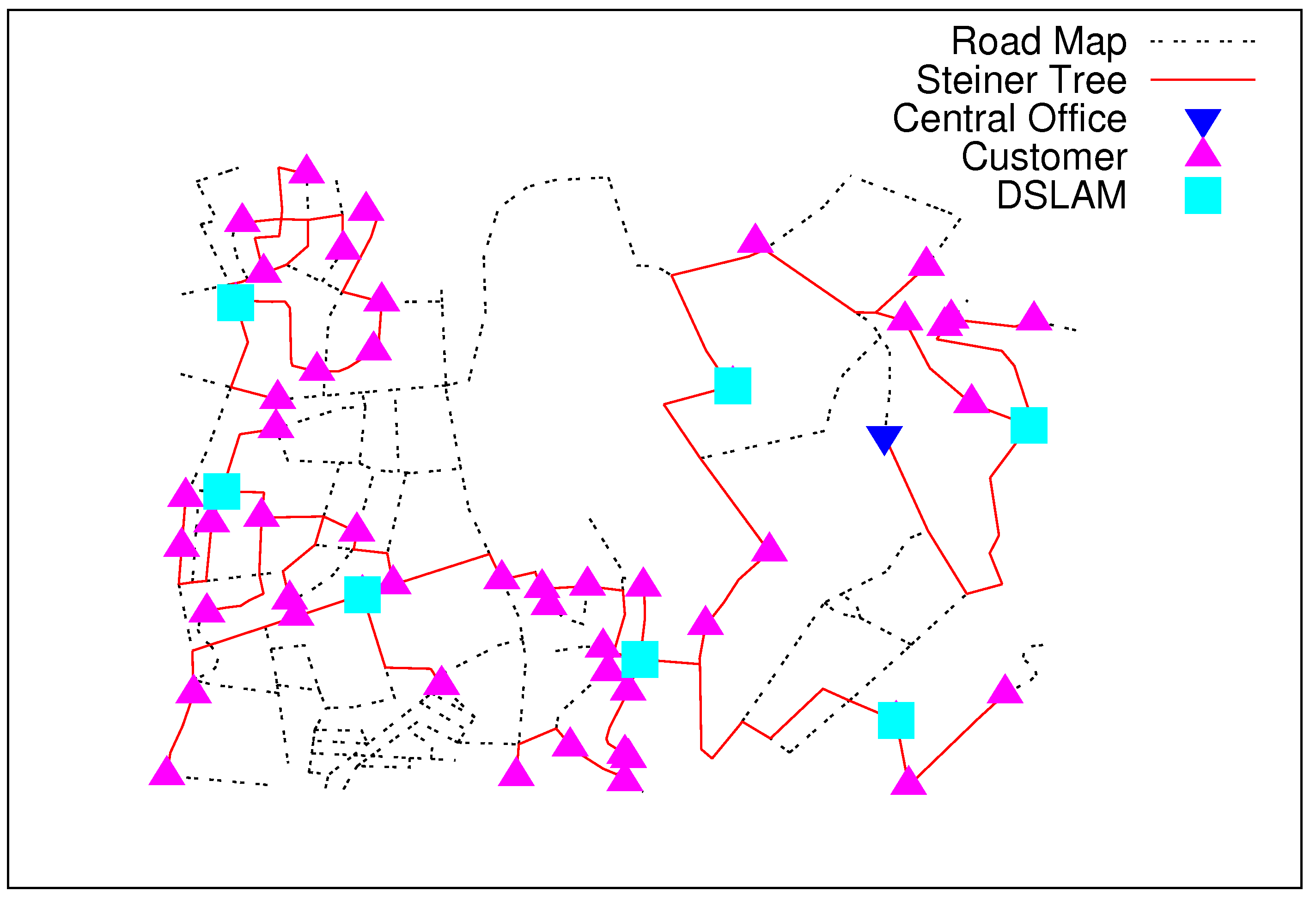}
            \label{fig:Steiner-map}}
        \hspace{0.05in}
        \subfloat[]{
            \includegraphics[width=1.6in]{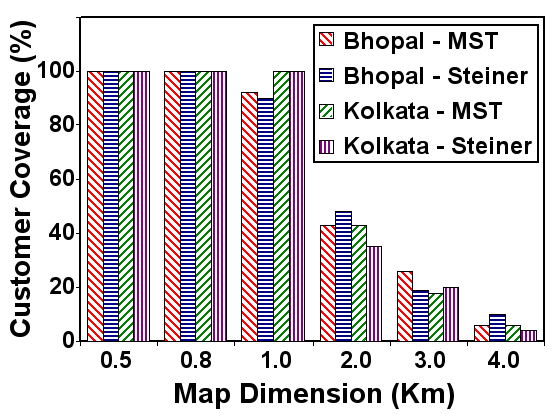}
            \label{fig:map-coverage}}
    \vspace*{-0.1in}
    \caption{A Steiner tree and remote DSLAM placement for a $2.0km \times 2.0km$ area of bhopal, and the portion of customers covered with copper loops from the node site when no remote DSLAMs are placed.}
    \label{fig:map}
    \end{center}
\end{figure}

Figure~\ref{fig:Steiner-map} shows a sample solution tree built over a $2.0km \times 2.0km$ area of Bhopal.
Figure~\ref{fig:map-coverage} shows the coverage of customers without using remote DSLAM units---as the area increases, fewer customers can be reached from the node site using loop lengths up to 1500m.

\begin{figure}[!h]
    \begin{center}
        \subfloat[]{
            \includegraphics[width=1.6in]{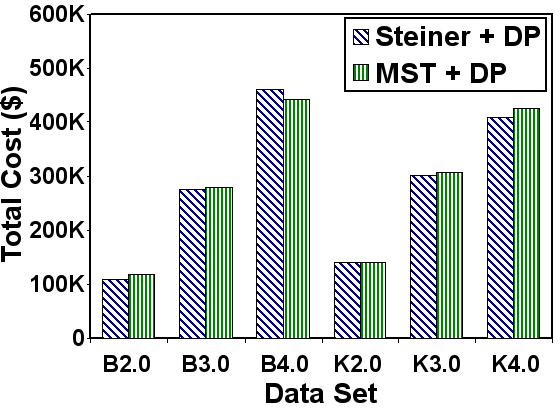}
            \label{fig:MST-Steiner}}
        \hspace{0.05in}
        \subfloat[]{
            \includegraphics[width=1.6in]{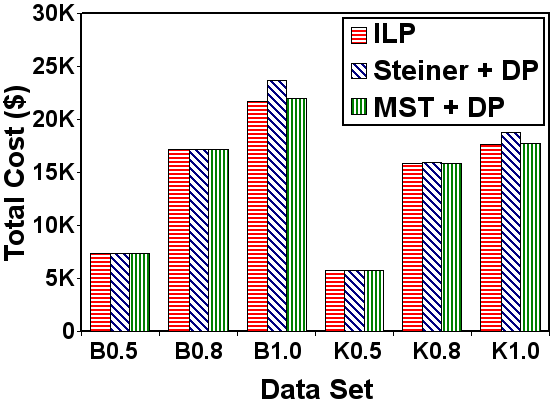}
            \label{fig:anything}}
    \vspace*{-0.1in}
    \caption{The relative performance of the Steiner and MST heuristic solutions for large datasets, and their performance compared to the optimal ILP for smaller datasets.}
    \label{fig:heuristic_cost}
    \end{center}
\end{figure}

 Figure~\ref{fig:MST-Steiner} shows the relative performance of our two heuristic-based approaches: Steiner and MST.  Both are very close, and we can simply run both and select the better solution for every dataset.  Figure~\ref{fig:anything} shows the performance of the heuristic approaches compared with the ILP for smaller datasets on were tractable for the ILP.  The best heuristic's performance is {\em very} close (within 1.5\%) of the optimal ILP solution.  We therefore restrict the remaining experiments on the larger datasets using the heuristic solution.

\begin{figure}[!h]
    \begin{center}
        \subfloat[Bhopal]{
            \includegraphics[width=1.6in]{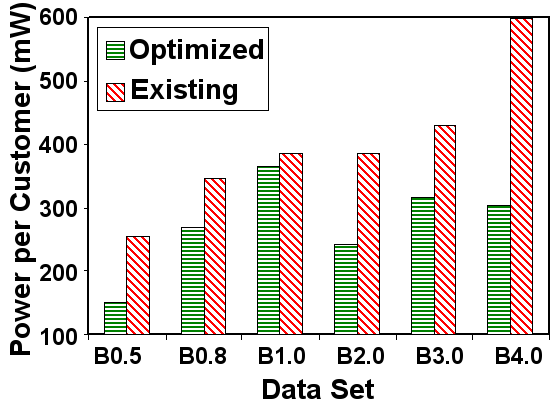}
            \label{fig:bhopal-power}}
        \hspace{0.05in}
        \subfloat[Kolkata]{
            \includegraphics[width=1.6in]{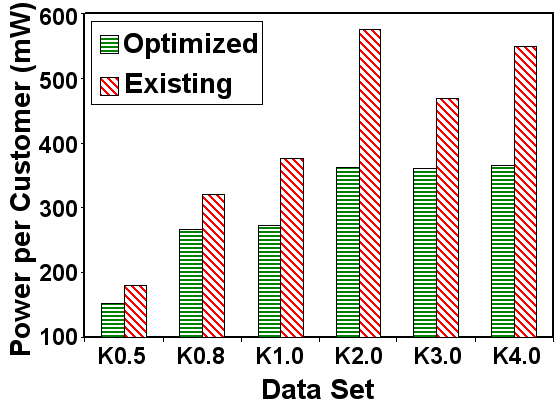}
            \label{fig:kolkata-power}}
    \vspace*{-0.1in}
    \caption{The reduced power per customer using the heuristic solution to place remote units.}
    \label{fig:heuristic_power}
    \end{center}
\end{figure}

\begin{figure}[!h]
    \begin{center}
        \subfloat[]{
\includegraphics[width=1.6in]{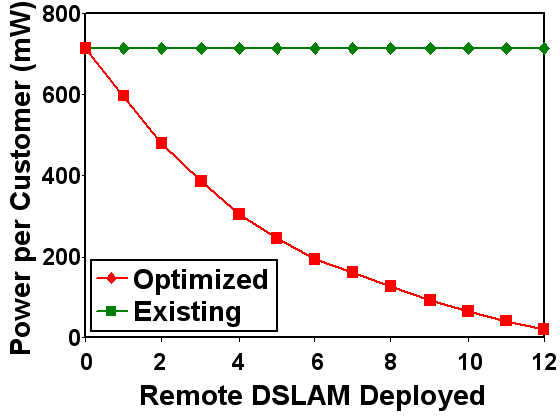}
            \label{fig:bhopal-power}}
        \hspace{0.05in}
        \subfloat[]{
            \includegraphics[width=1.6in]{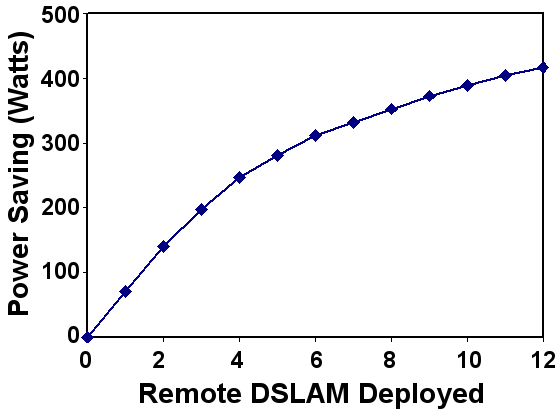}
            \label{fig:kolkata-power}}
    \vspace*{-0.1in}
    \caption{The reduced power per customer enabled as remote units are added to the network.}
    \label{fig:budget_bhopal}
    \end{center}
\end{figure}

Our greenfield solution increases coverage to 100\% by placing remote DSLAM units, while reducing the per-user power consumption, as shown in Figure~\ref{fig:heuristic_power}.   Finally, Figure~\ref{fig:budget_bhopal} shows the benefit of incrementally adding remote DSLAMs (implying additional expense) on the power consumption of the network.

\section{Conclusions}
Reducing the power consumption of DSL equipment has become an important research problem. Most approaches have focused on careful power management solutions to reduce the line driver power for individual lines without affecting line stability.  Our solution is complimentary and looks instead at how to design DSL access networks at an architectural level.   As far as we know, this is the first attempt to look at the problem from an access network-wide level. This paper formalizes the problem as a minimum-cost optimization problem, since high energy expenditure is driving many operators to look at  low-power access equipment.  By optimally placing remote DSLAM units between the access node site and the individual customers, we attempt to balance the cost of these units, laying fiber to them, and maintaining them with the savings in energy consumption and reaching additional set of customers.
The greenfield network layout problem is difficult to solve optimally, but our heuristic technique produces results that are very close.
Our approach can be used by operators to power-optimize their existing networks, or to design green-field deployments with a mix of fiber and copper.

\comment{
\section*{Acknowledgment}

The authors would like to thank Koen Hooghe for several useful data about DSL systems. He along with Mamoun Guenach and Jochen Maes also provided results of physical-layer simulations for copper loops.
}
\bibliographystyle{plain}

{\small
\bibliography{green1}
}

%
\end{document}